\documentclass[10pt,conference,twoside]{IEEEtran}
\usepackage{cite}
\usepackage[T1]{fontenc}
\usepackage{graphicx}

\usepackage{amssymb}
\usepackage{amsmath}
\usepackage{paralist}
\usepackage{subfigure}
\usepackage{booktabs} 
\usepackage{algorithm}
\usepackage{algorithmic}
\usepackage{amsthm}
\usepackage{multirow}
\usepackage{microtype}
\usepackage{tablefootnote}
\usepackage{dsfont}

\usepackage{amssymb}
\usepackage{amsfonts}
\usepackage{mathrsfs}
\usepackage{xspace}
\usepackage{bm}
\usepackage{upgreek}

\newcommand{\safemath}[2]{\newcommand{#1}{\ensuremath{#2}\xspace}}

\safemath{\bma}{\mathbf{a}}
\safemath{\bmb}{\mathbf{b}}
\safemath{\bmc}{\mathbf{c}}
\safemath{\bmd}{\mathbf{d}}
\safemath{\bme}{\mathbf{e}}
\safemath{\bmf}{\mathbf{f}}
\safemath{\bmg}{\mathbf{g}}
\safemath{\bmh}{\mathbf{h}}
\safemath{\bmi}{\mathbf{i}}
\safemath{\bmj}{\mathbf{j}}
\safemath{\bmk}{\mathbf{k}}
\safemath{\bml}{\mathbf{l}}
\safemath{\bmm}{\mathbf{m}}
\safemath{\bmn}{\mathbf{n}}
\safemath{\bmo}{\mathbf{o}}
\safemath{\bmp}{\mathbf{p}}
\safemath{\bmq}{\mathbf{q}}
\safemath{\bmr}{\mathbf{r}}
\safemath{\bms}{\mathbf{s}}
\safemath{\bmt}{\mathbf{t}}
\safemath{\bmu}{\mathbf{u}}
\safemath{\bmv}{\mathbf{v}}
\safemath{\bmw}{\mathbf{w}}
\safemath{\bmx}{\mathbf{x}}
\safemath{\bmy}{\mathbf{y}}
\safemath{\bmz}{\mathbf{z}}
\safemath{\bmzero}{\mathbf{0}}
\safemath{\bmone}{\mathbf{1}}

\bmdefine{\biad}{a}
\bmdefine{\bibd}{b}
\bmdefine{\bicd}{c}
\bmdefine{\bidd}{d}
\bmdefine{\bied}{e}
\bmdefine{\bifd}{f}
\bmdefine{\bigd}{g}
\bmdefine{\bihd}{h}
\bmdefine{\biid}{i}
\bmdefine{\bijd}{j}
\bmdefine{\bikd}{k}
\bmdefine{\bild}{l}
\bmdefine{\bimd}{m}
\bmdefine{\bind}{n}
\bmdefine{\biod}{o}
\bmdefine{\bipd}{p}
\bmdefine{\biqd}{q}
\bmdefine{\bird}{r}
\bmdefine{\bisd}{s}
\bmdefine{\bitd}{t}
\bmdefine{\biud}{u}
\bmdefine{\bivd}{v}
\bmdefine{\biwd}{w}
\bmdefine{\bixd}{x}
\bmdefine{\biyd}{y}
\bmdefine{\bizd}{z}

\bmdefine{\bixid}{\xi}
\bmdefine{\bilambdad}{\lambda}
\bmdefine{\bimud}{\mu}
\bmdefine{\bithetad}{\theta}
\bmdefine{\biphid}{\phi}
\bmdefine{\bideltad}{\delta}

\safemath{\bmia}{\biad}
\safemath{\bmib}{\bibd}
\safemath{\bmic}{\bicd}
\safemath{\bmid}{\bidd}
\safemath{\bmie}{\bied}
\safemath{\bmif}{\bifd}
\safemath{\bmig}{\bigd}
\safemath{\bmih}{\bihd}
\safemath{\bmii}{\biid}
\safemath{\bmij}{\bijd}
\safemath{\bmik}{\bikd}
\safemath{\bmil}{\bild}
\safemath{\bmim}{\bimd}
\safemath{\bmin}{\bind}
\safemath{\bmio}{\biod}
\safemath{\bmip}{\bipd}
\safemath{\bmiq}{\biqd}
\safemath{\bmir}{\bird}
\safemath{\bmis}{\bisd}
\safemath{\bmit}{\bitd}
\safemath{\bmiu}{\biud}
\safemath{\bmiv}{\bivd}
\safemath{\bmiw}{\biwd}
\safemath{\bmix}{\bixd}
\safemath{\bmiy}{\biyd}
\safemath{\bmiz}{\bizd}

\safemath{\bmxi}{\bixid}
\safemath{\bmlambda}{\bilambdad}
\safemath{\bmmu}{\bimud}
\safemath{\bmtheta}{\bithetad}
\safemath{\bmphi}{\biphid}
\safemath{\bmdelta}{\bideltad}

\safemath{\bA}{\mathbf{A}}
\safemath{\bB}{\mathbf{B}}
\safemath{\bC}{\mathbf{C}}
\safemath{\bD}{\mathbf{D}}
\safemath{\bE}{\mathbf{E}}
\safemath{\bF}{\mathbf{F}}
\safemath{\bG}{\mathbf{G}}
\safemath{\bH}{\mathbf{H}}
\safemath{\bI}{\mathbf{I}}
\safemath{\bJ}{\mathbf{J}}
\safemath{\bK}{\mathbf{K}}
\safemath{\bL}{\mathbf{L}}
\safemath{\bM}{\mathbf{M}}
\safemath{\bN}{\mathbf{N}}
\safemath{\bO}{\mathbf{O}}
\safemath{\bP}{\mathbf{P}}
\safemath{\bQ}{\mathbf{Q}}
\safemath{\bR}{\mathbf{R}}
\safemath{\bS}{\mathbf{S}}
\safemath{\bT}{\mathbf{T}}
\safemath{\bU}{\mathbf{U}}
\safemath{\bV}{\mathbf{V}}
\safemath{\bW}{\mathbf{W}}
\safemath{\bX}{\mathbf{X}}
\safemath{\bY}{\mathbf{Y}}
\safemath{\bZ}{\mathbf{Z}}

\safemath{\bZero}{\mathbf{0}}
\safemath{\bOne}{\mathbf{1}}
\safemath{\bDelta}{\mathbf{\Delta}}
\safemath{\bLambda}{\mathbf{\UpLambda}}
\safemath{\bPhi}{\mathbf{\Upphi}}
\safemath{\bSigma}{\mathbf{\Upsigma}}
\safemath{\bOmega}{\mathbf{\Upomega}}
\safemath{\bTheta}{\mathbf{\Uptheta}}

\bmdefine{\biAd}{A}
\bmdefine{\biBd}{B}
\bmdefine{\biCd}{C}
\bmdefine{\biDd}{D}
\bmdefine{\biEd}{E}
\bmdefine{\biFd}{F}
\bmdefine{\biGd}{G}
\bmdefine{\biHd}{H}
\bmdefine{\biId}{I}
\bmdefine{\biJd}{J}
\bmdefine{\biKd}{K}
\bmdefine{\biLd}{L}
\bmdefine{\biMd}{M}
\bmdefine{\biOd}{N}
\bmdefine{\biPd}{O}
\bmdefine{\biQd}{P}
\bmdefine{\biRd}{R}
\bmdefine{\biSd}{S}
\bmdefine{\biTd}{T}
\bmdefine{\biUd}{U}
\bmdefine{\biVd}{V}
\bmdefine{\biWd}{W}
\bmdefine{\biXd}{X}
\bmdefine{\biYd}{Y}
\bmdefine{\biZd}{Z}

\bmdefine{\biDelta}{\Delta}
\bmdefine{\biLambda}{\Lambda}
\bmdefine{\biPhi}{\Phi}
\bmdefine{\biSigma}{\Sigma}
\bmdefine{\biOmega}{\Omega}
\bmdefine{\biTheta}{\Theta}

\safemath{\bimA}{\biAd}
\safemath{\bimB}{\biBd}
\safemath{\bimC}{\biCd}
\safemath{\bimD}{\biDd}
\safemath{\bimE}{\biEd}
\safemath{\bimF}{\biFd}
\safemath{\bimG}{\biGd}
\safemath{\bimH}{\biHd}
\safemath{\bimI}{\biId}
\safemath{\bimJ}{\biJd}
\safemath{\bimK}{\biKd}
\safemath{\bimL}{\biLd}
\safemath{\bimM}{\biMd}
\safemath{\bimN}{\biNd}
\safemath{\bimO}{\biOd}
\safemath{\bimP}{\biPd}
\safemath{\bimQ}{\biQd}
\safemath{\bimR}{\biRd}
\safemath{\bimS}{\biSd}
\safemath{\bimT}{\biTd}
\safemath{\bimU}{\biUd}
\safemath{\bimV}{\biVd}
\safemath{\bimW}{\biWd}
\safemath{\bimX}{\biXd}
\safemath{\bimY}{\biYd}
\safemath{\bimZ}{\biZd}

\safemath{\bimDelta}{\biDelta}
\safemath{\bimLambda}{\biLambda}
\safemath{\bimPhi}{\biPhi}
\safemath{\bimSigma}{\biSigma}
\safemath{\bimOmega}{\biOmega}
\safemath{\bimTheta}{\biTheta}

\safemath{\setA}{\mathcal{A}}
\safemath{\setB}{\mathcal{B}}
\safemath{\setC}{\mathcal{C}}
\safemath{\setD}{\mathcal{D}}
\safemath{\setE}{\mathcal{E}}
\safemath{\setF}{\mathcal{F}}
\safemath{\setG}{\mathcal{G}}
\safemath{\setH}{\mathcal{H}}
\safemath{\setI}{\mathcal{I}}
\safemath{\setJ}{\mathcal{J}}
\safemath{\setK}{\mathcal{K}}
\safemath{\setL}{\mathcal{L}}
\safemath{\setM}{\mathcal{M}}
\safemath{\setN}{\mathcal{N}}
\safemath{\setO}{\mathcal{O}}
\safemath{\setP}{\mathcal{P}}
\safemath{\setQ}{\mathcal{Q}}
\safemath{\setR}{\mathcal{R}}
\safemath{\setS}{\mathcal{S}}
\safemath{\setT}{\mathcal{T}}
\safemath{\setU}{\mathcal{U}}
\safemath{\setV}{\mathcal{V}}
\safemath{\setW}{\mathcal{W}}
\safemath{\setX}{\mathcal{X}}
\safemath{\setY}{\mathcal{Y}}
\safemath{\setZ}{\mathcal{Z}}
\safemath{\emptySet}{\varnothing}

\safemath{\colA}{\mathscr{A}}
\safemath{\colB}{\mathscr{B}}
\safemath{\colC}{\mathscr{C}}
\safemath{\colD}{\mathscr{D}}
\safemath{\colE}{\mathscr{E}}
\safemath{\colF}{\mathscr{F}}
\safemath{\colG}{\mathscr{G}}
\safemath{\colH}{\mathscr{H}}
\safemath{\colI}{\mathscr{I}}
\safemath{\colJ}{\mathscr{J}}
\safemath{\colK}{\mathscr{K}}
\safemath{\colL}{\mathscr{L}}
\safemath{\colM}{\mathscr{M}}
\safemath{\colN}{\mathscr{N}}
\safemath{\colO}{\mathscr{O}}
\safemath{\colP}{\mathscr{P}}
\safemath{\colQ}{\mathscr{Q}}
\safemath{\colR}{\mathscr{R}}
\safemath{\colS}{\mathscr{S}}
\safemath{\colT}{\mathscr{T}}
\safemath{\colU}{\mathscr{U}}
\safemath{\colV}{\mathscr{V}}
\safemath{\colW}{\mathscr{W}}
\safemath{\colX}{\mathscr{X}}
\safemath{\colY}{\mathscr{Y}}
\safemath{\colZ}{\mathscr{Z}}

\safemath{\opA}{\mathbb{A}}
\safemath{\opB}{\mathbb{B}}
\safemath{\opC}{\mathbb{C}}
\safemath{\opD}{\mathbb{D}}
\safemath{\opE}{\mathbb{E}}
\safemath{\opF}{\mathbb{F}}
\safemath{\opG}{\mathbb{G}}
\safemath{\opH}{\mathbb{H}}
\safemath{\opI}{\mathbb{I}}
\safemath{\opJ}{\mathbb{J}}
\safemath{\opK}{\mathbb{K}}
\safemath{\opL}{\mathbb{L}}
\safemath{\opM}{\mathbb{M}}
\safemath{\opN}{\mathbb{N}}
\safemath{\opO}{\mathbb{O}}
\safemath{\opP}{\mathbb{P}}
\safemath{\opQ}{\mathbb{Q}}
\safemath{\opR}{\mathbb{R}}
\safemath{\opS}{\mathbb{S}}
\safemath{\opT}{\mathbb{T}}
\safemath{\opU}{\mathbb{U}}
\safemath{\opV}{\mathbb{V}}
\safemath{\opW}{\mathbb{W}}
\safemath{\opX}{\mathbb{X}}
\safemath{\opY}{\mathbb{Y}}
\safemath{\opZ}{\mathbb{Z}}
\safemath{\opZero}{\mathbb{O}}
\safemath{\identityop}{\opI}

\safemath{\veca}{\bma}
\safemath{\vecb}{\bmb}
\safemath{\vecc}{\bmc}
\safemath{\vecd}{\bmd}
\safemath{\vece}{\bme}
\safemath{\vecf}{\bmf}
\safemath{\vecg}{\bmg}
\safemath{\vech}{\bmh}
\safemath{\veci}{\bmi}
\safemath{\vecj}{\bmj}
\safemath{\veck}{\bmk}
\safemath{\vecl}{\bml}
\safemath{\vecm}{\bmm}
\safemath{\vecn}{\bmn}
\safemath{\veco}{\bmo}
\safemath{\vecp}{\bmp}
\safemath{\vecq}{\bmq}
\safemath{\vecr}{\bmr}
\safemath{\vecs}{\bms}
\safemath{\vect}{\bmt}
\safemath{\vecu}{\bmu}
\safemath{\vecv}{\bmv}
\safemath{\vecw}{\bmw}
\safemath{\vecx}{\bmx}
\safemath{\vecy}{\bmy}
\safemath{\vecz}{\bmz}

\safemath{\veczero}{\bmzero}
\safemath{\vecone}{\bmone}
\safemath{\vecxi}{\bmxi}
\safemath{\veclambda}{\bmlambda}
\safemath{\vecmu}{\bmmu}
\safemath{\vectheta}{\bmtheta}
\safemath{\vecphi}{\bmphi}
\safemath{\vecdelta}{\bmdelta}

\safemath{\matA}{\bA}
\safemath{\matB}{\bB}
\safemath{\matC}{\bC}
\safemath{\matD}{\bD}
\safemath{\matE}{\bE}
\safemath{\matF}{\bF}
\safemath{\matG}{\bG}
\safemath{\matH}{\bH}
\safemath{\matI}{\bI}
\safemath{\matJ}{\bJ}
\safemath{\matK}{\bK}
\safemath{\matL}{\bL}
\safemath{\matM}{\bM}
\safemath{\matN}{\bN}
\safemath{\matO}{\bO}
\safemath{\matP}{\bP}
\safemath{\matQ}{\bQ}
\safemath{\matR}{\bR}
\safemath{\matS}{\bS}
\safemath{\matT}{\bT}
\safemath{\matU}{\bU}
\safemath{\matV}{\bV}
\safemath{\matW}{\bW}
\safemath{\matX}{\bX}
\safemath{\matY}{\bY}
\safemath{\matZ}{\bZ}
\safemath{\matzero}{\bmzero}

\safemath{\matDelta}{\bDelta}
\safemath{\matLambda}{\bLambda}
\safemath{\matPhi}{\bPhi}
\safemath{\matSigma}{\bSigma}
\safemath{\matOmega}{\bOmega}
\safemath{\matTheta}{\bTheta}

\safemath{\matidentity}{\matI}
\safemath{\matone}{\matO}

\safemath{\rnda}{A}
\safemath{\rndb}{B}
\safemath{\rndc}{C}
\safemath{\rndd}{D}
\safemath{\rnde}{E}
\safemath{\rndf}{F}
\safemath{\rndg}{G}
\safemath{\rndh}{H}
\safemath{\rndi}{I}
\safemath{\rndj}{J}
\safemath{\rndk}{K}
\safemath{\rndl}{L}
\safemath{\rndm}{M}
\safemath{\rndn}{N}
\safemath{\rndo}{O}
\safemath{\rndp}{P}
\safemath{\rndq}{Q}
\safemath{\rndr}{R}
\safemath{\rnds}{S}
\safemath{\rndt}{T}
\safemath{\rndu}{U}
\safemath{\rndv}{V}
\safemath{\rndw}{W}
\safemath{\rndx}{X}
\safemath{\rndy}{Y}
\safemath{\rndz}{Z}

\safemath{\rveca}{\bimA}
\safemath{\rvecb}{\bimB}
\safemath{\rvecc}{\bimC}
\safemath{\rvecd}{\bimD}
\safemath{\rvece}{\bimE}
\safemath{\rvecf}{\bimF}
\safemath{\rvecg}{\bimG}
\safemath{\rvech}{\bimH}
\safemath{\rveci}{\bimI}
\safemath{\rvecj}{\bimJ}
\safemath{\rveck}{\bimK}
\safemath{\rvecl}{\bimL}
\safemath{\rvecm}{\bimM}
\safemath{\rvecn}{\bimN}
\safemath{\rveco}{\bomO}
\safemath{\rvecp}{\bimP}
\safemath{\rvecq}{\bimQ}
\safemath{\rvecr}{\bimR}
\safemath{\rvecs}{\bimS}
\safemath{\rvect}{\bimT}
\safemath{\rvecu}{\bimU}
\safemath{\rvecv}{\bimV}
\safemath{\rvecw}{\bimW}
\safemath{\rvecx}{\bimX}
\safemath{\rvecy}{\bimY}
\safemath{\rvecz}{\bimZ}

\safemath{\rvecxi}{\bmxi}
\safemath{\rveclambda}{\bmlambda}
\safemath{\rvecmu}{\bmmu}
\safemath{\rvectheta}{\bmtheta}
\safemath{\rvecphi}{\bmphi}

\safemath{\rmatA}{\bimA}
\safemath{\rmatB}{\bimB}
\safemath{\rmatC}{\bimC}
\safemath{\rmatD}{\bimD}
\safemath{\rmatE}{\bimE}
\safemath{\rmatF}{\bimF}
\safemath{\rmatG}{\bimG}
\safemath{\rmatH}{\bimH}
\safemath{\rmatI}{\bimI}
\safemath{\rmatJ}{\bimJ}
\safemath{\rmatK}{\bimK}
\safemath{\rmatL}{\bimL}
\safemath{\rmatM}{\bimM}
\safemath{\rmatN}{\bimN}
\safemath{\rmatO}{\bimO}
\safemath{\rmatP}{\bimP}
\safemath{\rmatQ}{\bimQ}
\safemath{\rmatR}{\bimR}
\safemath{\rmatS}{\bimS}
\safemath{\rmatT}{\bimT}
\safemath{\rmatU}{\bimU}
\safemath{\rmatV}{\bimV}
\safemath{\rmatW}{\bimW}
\safemath{\rmatX}{\bimX}
\safemath{\rmatY}{\bimY}
\safemath{\rmatZ}{\bimZ}

\safemath{\rmatDelta}{\bimDelta}
\safemath{\rmatLambda}{\bimLambda}
\safemath{\rmatPhi}{\bimPhi}
\safemath{\rmatSigma}{\bimSigma}
\safemath{\rmatOmega}{\bimOmega}
\safemath{\rmatTheta}{\bimTheta}
 
\usepackage{amssymb}
\usepackage{amsfonts}
\usepackage{mathrsfs}
\usepackage{xspace}
\usepackage{bm}
\usepackage{fancyref}
\usepackage{textcomp}

\usepackage{multirow}
\usepackage{stmaryrd}

\newenvironment{textbmatrix}{	\setlength{\arraycolsep}{2.5pt}%
								\big[\begin{matrix}}{\end{matrix}\big]%
								\raisebox{0.08ex}{\vphantom{M}}}

\def\be{\begin{equation}}
\def\ee{\end{equation}}
\def\een{\nonumber \end{equation}}
\def\mat{\begin{bmatrix}}
\def\emat{\end{bmatrix}}
\def\btm{\begin{textbmatrix}}
\def\etm{\end{textbmatrix}}

\def\ba#1\ea{\begin{align}#1\end{align}}
\def\bas#1\eas{\begin{align*}#1\end{align*}}
\def\bs#1\es{\begin{split}#1\end{split}}
\def\bg#1\eg{\begin{gather}#1\end{gather}}
\def\bml#1\eml{\begin{multline}#1\end{multline}}
\def\bi#1\ei{\begin{itemize}#1\end{itemize}}

\newcommand{\lefto}{\mathopen{}\left}

\DeclareMathOperator{\sign}{sign}			
\DeclareMathOperator*{\argmin}{arg\;min}		
\DeclareMathOperator*{\argmax}{arg\;max}		
\DeclareMathOperator{\Exop}{\opE}			
\DeclareMathOperator{\Varop}{\opV\!\mathrm{ar}} 

\newcommand{\abs}[1]{\lefto\lvert#1\right\rvert}		

\safemath{\dirac}{\delta}					
\safemath{\krond}{\dirac}					

\safemath{\upto}{\uparrow}
\safemath{\downto}{\downarrow}
\safemath{\iu}{j}							
\safemath{\ev}{\lambda}						
\safemath{\hilseqspace}{l^{2}}				
\newcommand{\banachfunspace}[1]{\setL^{#1}}	
\safemath{\hilfunspace}{\banachfunspace{2}}	

\safemath{\SNR}{\textit{SNR}} 				
\safemath{\PAR}{\textit{PAR}} 				
\safemath{\No}{N_0}							
\safemath{\Es}{E_s}							
\safemath{\Eb}{E_b}							
\safemath{\EbNo}{\frac{\Eb}{\No}}
\safemath{\EsNo}{\frac{\Es}{\No}}

\DeclareMathOperator{\CHop}{\ensuremath{\opH}} 
\safemath{\tvir}{\rndh_{\CHop}}				
\safemath{\tvtf}{\rndl_{\CHop}}				
\safemath{\spf}{\rnds_{\CHop}}				
\safemath{\bff}{H_{\CHop}}					

\safemath{\ircf}{r_{h}}						
\safemath{\tftvcf}{r_{s}}					
\safemath{\tfcf}{r_{l}}						
\safemath{\bfcf}{r_{H}}						

\safemath{\tcorr}{c_h}						
\safemath{\scf}{c_{s}}						
\safemath{\tfcorr}{c_{l}}					
\safemath{\fcorr}{c_{H}}						

\safemath{\mi}{I}							
\safemath{\capacity}{C}						

\safemath{\normal}{\mathcal{N}}			
\safemath{\jpg}{\mathcal{CN}}			
\safemath{\mchain}{\leftrightarrow}		

\safemath{\dB}{\,\mathrm{dB}}
\safemath{\dBm}{\,\mathrm{dBm}}
\safemath{\Hz}{\,\mathrm{Hz}}
\safemath{\kHz}{\,\mathrm{kHz}}
\safemath{\MHz}{\,\mathrm{MHz}}
\safemath{\GHz}{\,\mathrm{GHz}}
\safemath{\s}{\,\mathrm{s}}
\safemath{\ms}{\,\mathrm{ms}}
\safemath{\mus}{\,\mathrm{\text{\textmu}s}}
\safemath{\ns}{\,\mathrm{ns}}
\safemath{\ps}{\,\mathrm{ps}}
\safemath{\meter}{\,\mathrm{m}}
\safemath{\mm}{\,\mathrm{mm}}
\safemath{\cm}{\,\mathrm{cm}}
\safemath{\m}{\,\mathrm{m}}
\safemath{\W}{\,\mathrm{W}}
\safemath{\mW}{\, \mathrm{mW}}
\safemath{\J}{\,\mathrm{J}}
\safemath{\K}{\,\mathrm{K}}
\safemath{\bit}{\,\mathrm{bit}}
\safemath{\nat}{\,\mathrm{nat}}

\safemath{\define}{\triangleq}			

\safemath{\equivalent}{\sim}
\safemath{\distas}{\sim}					
\safemath{\sdiff}{\Delta}				

\safemath{\reals}{\mathbb{R}}
\safemath{\positivereals}{\reals_{+}}
\safemath{\integers}{\mathbb{Z}}
\safemath{\posint}{\integers_{+}}
\safemath{\naturals}{\mathbb{N}}
\safemath{\posnaturals}{\naturals_{+}}
\safemath{\complexset}{\mathbb{C}}
\safemath{\rationals}{\mathbb{Q}}

\newcommand*{\fancyrefapplabelprefix}{app}		
\newcommand*{\fancyrefthmlabelprefix}{thm}		
\newcommand*{\fancyreflemlabelprefix}{lem}		
\newcommand*{\fancyrefcorlabelprefix}{cor}		
\newcommand*{\fancyrefdeflabelprefix}{def}		
\newcommand*{\fancyrefproplabelprefix}{prop}		
\newcommand*{\fancyrefexmpllabelprefix}{exmpl}
\newcommand*{\fancyrefalglabelprefix}{alg}		
\newcommand*{\fancyreftbllabelprefix}{tbl}		
\newcommand*{\fancyrefremlabelprefix}{rem}		

\frefformat{vario}{\fancyrefseclabelprefix}{Section~#1}
\frefformat{vario}{\fancyrefthmlabelprefix}{Theorem~#1}
\frefformat{vario}{\fancyreftbllabelprefix}{Table~#1}
\frefformat{vario}{\fancyreflemlabelprefix}{Lemma~#1}
\frefformat{vario}{\fancyrefcorlabelprefix}{Corollary~#1}
\frefformat{vario}{\fancyrefdeflabelprefix}{Definition~#1}
\frefformat{vario}{\fancyreffiglabelprefix}{Fig.~#1}
\frefformat{vario}{\fancyrefapplabelprefix}{Appendix~#1}
\frefformat{vario}{\fancyrefeqlabelprefix}{(#1)}
\frefformat{vario}{\fancyrefproplabelprefix}{Proposition~#1}
\frefformat{vario}{\fancyrefexmpllabelprefix}{Example~#1}
\frefformat{vario}{\fancyrefalglabelprefix}{Algorithm~#1}
\frefformat{vario}{\fancyrefremlabelprefix}{Remark~#1}

\newtheorem{thm}{Theorem}

\newtheorem{defi}{Definition}
\newtheorem{lem}[thm]{Lemma}

 \newtheorem{rem}{Remark}

\safemath{\dictab}{[\,\dicta\,\,\dictb\,]}

\safemath{\ysig}{\bmy}
\safemath{\ysighat}{\hat{\ysig}}
\safemath{\ysigdim}{M}
\safemath{\xsig}{\bmx}
\safemath{\xsigdim}{N}
\safemath{\nx}{n_x}
\safemath{\zsig}{\bmz}
\safemath{\zsigdim}{\ysigdim}
\safemath{\rsig}{\bmr}
\safemath{\Adict}{\bA}
\safemath{\Adicttilde}{\widetilde{\Adict}}
\safemath{\Adictdim}{\outputdim\times\xsigdim}
\safemath{\avec}{\bma}
\safemath{\avectilde}{\tilde{\avec}}
\safemath{\Bdict}{\bB}
\safemath{\Bdicttilde}{\widetilde{\Bdict}}
\safemath{\Cdict}{\bC}
\safemath{\cvec}{\bmc}
\safemath{\Ddict}{\bD}
\safemath{\Ddictdim}{\ysigdim\times\xsigdim}
\safemath{\dvec}{\bmd}
\safemath{\Ddicttilde}{\widetilde{\bD}}
\safemath{\Bonb}{\bB}
\safemath{\bvec}{\bmb}
\safemath{\Bonbdim}{\ysigdim\times\ysigdim}
\safemath{\noise}{\bmn}
\safemath{\noisedim}{\ysigim}
\safemath{\err}{\bme}
\safemath{\errdim}{\ysigdim}
\safemath{\errset}{\setE}
\safemath{\nerr}{n_e}
\safemath{\delop}{\bP_\errset}
\safemath{\delopc}{\bP_{{\errset}^c}}

\safemath{\cplxi}{\imath}
\safemath{\cplxj}{\jmath}

\safemath{\dict}{\matD}
\safemath{\inputdim}{N}		
\safemath{\outputdim}{M}		
\safemath{\sparsity}{S}	
\safemath{\inputdimA}{{N_a}}	
\safemath{\inputdimB}{{N_b}}	
\safemath{\elemA}{{n_a}}	
\safemath{\elemB}{{n_b}}	
\safemath{\resA}{\matR_a}	
\safemath{\resB}{\matR_b}	
\safemath{\subD}{\matS} 
\safemath{\subA}{\matS_a} 
\safemath{\subB}{\matS_b} 
\safemath{\dicta}{\matA} 	
\safemath{\dictb}{\matB} 	
\safemath{\hollowS}{H}
\safemath{\hollowA}{H_a}
\safemath{\hollowB}{H_b}
\safemath{\cross}{Z}
\safemath{\coh}{\mu_d}			
\safemath{\coha}{\mu_a}			
\safemath{\cohb}{\mu_b}			
\safemath{\mubs}{\nu}	
\safemath{\cohm}{\mu_m} 
\safemath{\dictset}{\setD}	
\safemath{\dictsetp}{\dictset(\coh,\coha,\cohb)}	
\safemath{\dictsetgen}{\dictset_\text{gen}}
\safemath{\dictsetgenp}{\dictsetgen(\coh)}
\safemath{\dictsetonb}{\dictset_\text{onb}}
\safemath{\dictsetonbp}{\dictsetonb(\coh)}

\safemath{\leftside}{U}
\safemath{\rightsideA}{R_a}
\safemath{\rightsideB}{R_b}

\safemath{\indexS}{\setI_S} 

\safemath{\na}{n_a}			
\safemath{\nb}{n_b}			
\safemath{\coeffa}{p_i}	
\safemath{\coeffb}{q_j}	
\safemath{\seta}{\setP}		
\safemath{\setb}{\setQ}     
\safemath{\setw}{\setW}	
\safemath{\setz}{\setZ}	
\safemath{\cola}{\veca}		
\safemath{\colb}{\vecb}		
\safemath{\cold}{\vecd}		
\safemath{\inputvec}{\vecx} 	
\safemath{\error}{\vece}	
\safemath{\noiseout}{\vecz} 	
\safemath{\inputvecel}{x}
\safemath{\inputveca}{\vecx_a}
\safemath{\inputvecb}{\vecx_b}
\safemath{\outputvec}{\vecy}	
\safemath{\lambdamin}{\lambda_{\mathrm{min}}}

\safemath{\elltwo}{\ell_2}
\safemath{\ellone}{\ell_1}
\safemath{\ellzero}{\ell_0}
\safemath{\ellinf}{\ell_\infty}
\safemath{\ellinftilde}{\ell_{\widetilde\infty}}
\safemath{\licard}{Z(\coh,\coha,\cohb)}
\safemath{\xsol}{\hat{x}}
\safemath{\xbord}{x_b}		
\safemath{\xstat}{x_s}		
\safemath{\xstatLone}{\tilde{x}_s}
\safemath{\order}{\mathcal{O}} 
\safemath{\scales}{\Theta} 
\safemath{\ones}{\mathbf{1}} 
\safemath{\zeroes}{\mathbf{0}} 
\safemath{\thlone}{\kappa(\coh,\cohb)} 
\safemath{\constoneA}{\delta} 
\safemath{\constoneB}{\epsilon} 
\safemath{\nlarge}{L}				   
\safemath{\sumlarge}{S_\nlarge}
\safemath{\maxlarger}{P_\nlarge}	   
\safemath{\Pzero}{\textrm{P0}}	
\safemath{\Pone}{\textrm{P1}}
\safemath{\vecfir}{\vecw}			 
\safemath{\vecsec}{\vecz}
\safemath{\elvecfir}{w}              
\safemath{\elvecsec}{z}				 
\safemath{\nlargefir}{n}
\safemath{\normout}{\gamma}
\safemath{\auxfun}{h}
\safemath{\supp}{\textrm{supp}}

\safemath{\indexa}{\ell}
\safemath{\indexb}{r}
\safemath{\indexc}{i}
\safemath{\indexd}{j}

\safemath{\project}{P}

\usepackage{geometry}
\usepackage[usenames,dvipsnames]{color}

\newcommand{\xrv}[1]{X#1}

\newcommand{\resid}[1]{\bmr#1}

\safemath{\NT}{N_\textnormal{T}}
\safemath{\tmax}{{t_\textnormal{max}}}
\safemath{\LAMA}{\textrm{LAMA}}
\safemath{\MRT}{\textrm{MRT}}
\safemath{\betamax}{\beta^\textnormal{max}}
\safemath{\betamin}{\beta^\textnormal{min}}

\safemath{\Nomin}{\No^\textnormal{min}(\beta)}
\safemath{\Nomax}{\No^\textnormal{max}(\beta)}

\safemath{\MAP}{\textnormal{MAP}}
\safemath{\IO}{\textnormal{IO}}
\safemath{\Opt}{\textnormal{IO}}
\safemath{\JO}{\textnormal{JO}}
\safemath{\Nopost}{N_{0}^\textnormal{post}}
\safemath{\MT}{{M_\textnormal{T}}}
\safemath{\MR}{{M_\textnormal{R}}}
\safemath{\Tran}{\textnormal{T}}
\safemath{\Herm}{\textnormal{H}}
\safemath{\row}{\textnormal{r}}
\safemath{\col}{\textnormal{c}}

\safemath{\dd}{\textnormal{d}}

\IEEEoverridecommandlockouts 

\begin{document}

\title{Optimal Large-MIMO Data Detection \\ with Transmit Impairments}

\author{Ramina Ghods, Charles Jeon, Arian Maleki, and Christoph Studer \\[0.1cm]
\thanks{R.~Ghods, C.~Jeon, and C.~Studer are with the School of Electrical and Computer Engineering, Cornell University, Ithaca, NY; e-mail: {rghods@csl.cornell.edu}, {jeon@csl.cornell.edu}, {studer@cornell.edu}.}
\thanks{A. Maleki is with Department of Statistics at Columbia University, New York City, NY; e-mail: {arian@stat.columbia.edu}.}
\thanks{The work of R. Ghods, C. Jeon, and C. Studer was supported in part by Xilinx Inc., and by the US National Science Foundation under grants ECCS-1408006 and CCF-1535897.}
\thanks{The work of A. Maleki was supported by the US National Science Foundation under grant CCF-1420328.} 
}
\maketitle

\begin{abstract}
Real-world transceiver designs for multiple-input multiple-output (MIMO) wireless communication systems are affected by a number of hardware impairments that already appear at the transmit side, such as amplifier non-linearities, quantization artifacts, and phase noise. While such transmit-side impairments are routinely ignored in the data-detection literature, they often limit reliable communication in practical systems. In this paper, we present a novel data-detection algorithm, referred to as \underline{la}rge-\underline{M}IMO \underline{a}pproximate message passing with transmit \underline{i}mpairments (short LAMA-I), which takes into account a broad range of transmit-side impairments in wireless systems with a large number of transmit and receive antennas. We provide conditions in the large-system limit for which LAMA-I achieves the  error-rate performance of the individually-optimal (IO) data detector. We furthermore demonstrate that LAMA-I achieves near-IO performance at low computational complexity in realistic, finite dimensional large-MIMO systems.
\end{abstract}



\section{Introduction}\label{sec:intro}

Practical transceiver implementations for wireless communication systems suffer from a number of radio-frequency~(RF) hardware impairments that already occur at the transmit side, including (but not limited to) amplifier non-linearities, quantization artifacts, and phase noise~\cite{Studer_Tx_OFDM,studer2011system,schenk2008rf,schenk2005performance,goransson2008effect,suzuki2008transmitter,suzuki2009practical,gonzalez2011impact,gonzalez2011transmit,bjornson2013capacity,zhang2014mimo}.
This paper deals with optimal data detection in the presence of such impairments for large (multi-user) multiple-input multiple-output (MIMO) wireless systems with a large number of antenna elements at (possibly) both ends of the wireless link~\cite{rusek2013scaling,marzetta2010noncooperative}.
In particular, we consider the problem of estimating the $\MT$-dimensional transmit data vector $\vecs\in\setO^\MT$, where~$\setO$ is a finite constellation set (e.g., QAM or PSK), observed from the following (impaired) MIMO input-output relation~\cite{Studer_Tx_OFDM,studer2011system}:
\begin{align}\label{eq:TNproblem}
\bmy = \bH(\bms+\bme) + \bmn.
\end{align}
Here,  the vector $\vecy\in\complexset^\MR$ corresponds to the received signal,  the matrix $\bH\in\complexset^{\MR\times\MT}$ represents the MIMO channel, the vector $\vece\in\complexset^{\MT}$ models transmit impairments, and the vector $\bmn\in\complexset^\MR$ corresponds to receive noise; the number of receive and transmit antennas is denoted by $\MR$ and~$\MT$, respectively. 

\subsection{Contributions}
\label{sec:contributions}

We build upon our previous results in \cite{jgms2015conf} and develop  a novel, computationally efficient data detection algorithm for the model \eqref{eq:TNproblem}, referred to as LAMA-I (short for \underline{la}rge-\underline{M}IMO \underline{a}pproximate message passing with transmit \underline{i}mpairments). 
%
%
We provide conditions for which \mbox{LAMA-I} achieves the error-rate performance of the individually optimal (IO) data-detector, which solves the following optimization problem:
\begin{align}\label{eq:IOproblem}
\hat{s}_\ell^\Opt=\argmin_{\tilde s_\ell \in \setO} \, \mathbb{P} \!\left( \tilde s_\ell \neq s_\ell \right)\!.
\end{align}
In words, \mbox{\mbox{LAMA-I}} aims at minimizing the per-user symbol-error probability~\cite{donoho2009,andreaGMCS}.
Assuming $p(\vecs)=\prod_{i=1}^{\MT}p(s_i)$ and i.i.d.\ circularly-symmetric complex Gaussian noise with variance $\No$ per complex entry of the noise vector $\vecn$, we define the \emph{effective transmit signal} $\vecx\in\complexset^\MT$ as $\bmx=\bms+\bme$ with the transmit-impairment distribution $p(\bmx|\bms)=\prod_{\ell=1}^{\MT}p(x_\ell|s_\ell)$. 
Besides user-wise independence, we do not impose any conditions on the statistics of the transmit impairments---this allows us to model a broad range of transmit-side impairments, including hardware non-idealities that exhibit statistical dependence between impairments and the data symbols, as well as deterministic effects (e.g., non-linearities).
%

Our optimality conditions are derived via the state-evolution (SE) framework \cite{donoho2009,andreaGMCS} of approximate message passing (AMP)\cite{Maleki2010phd,DMM10a,DMM10b} and for the asymptotic setting, i.e., the so-called large-system limit. Specifically, we fix the \emph{system ratio} $\beta=\MT/\MR$ and let $\MT\to\infty$, and assume that the entries of $\bH$ are i.i.d.\ circularly-symmetric complex Gaussian with variance $1/\MR$ per complex entry. 
To demonstrate the efficacy of \mbox{LAMA-I} in practice, we provide error-rate simulation results in finite-dimensional large-MIMO systems. 

\begin{figure*}
\centering
\subfigure[$128$ BS antennas and $8$ users.]{\includegraphics[width=0.98\columnwidth]{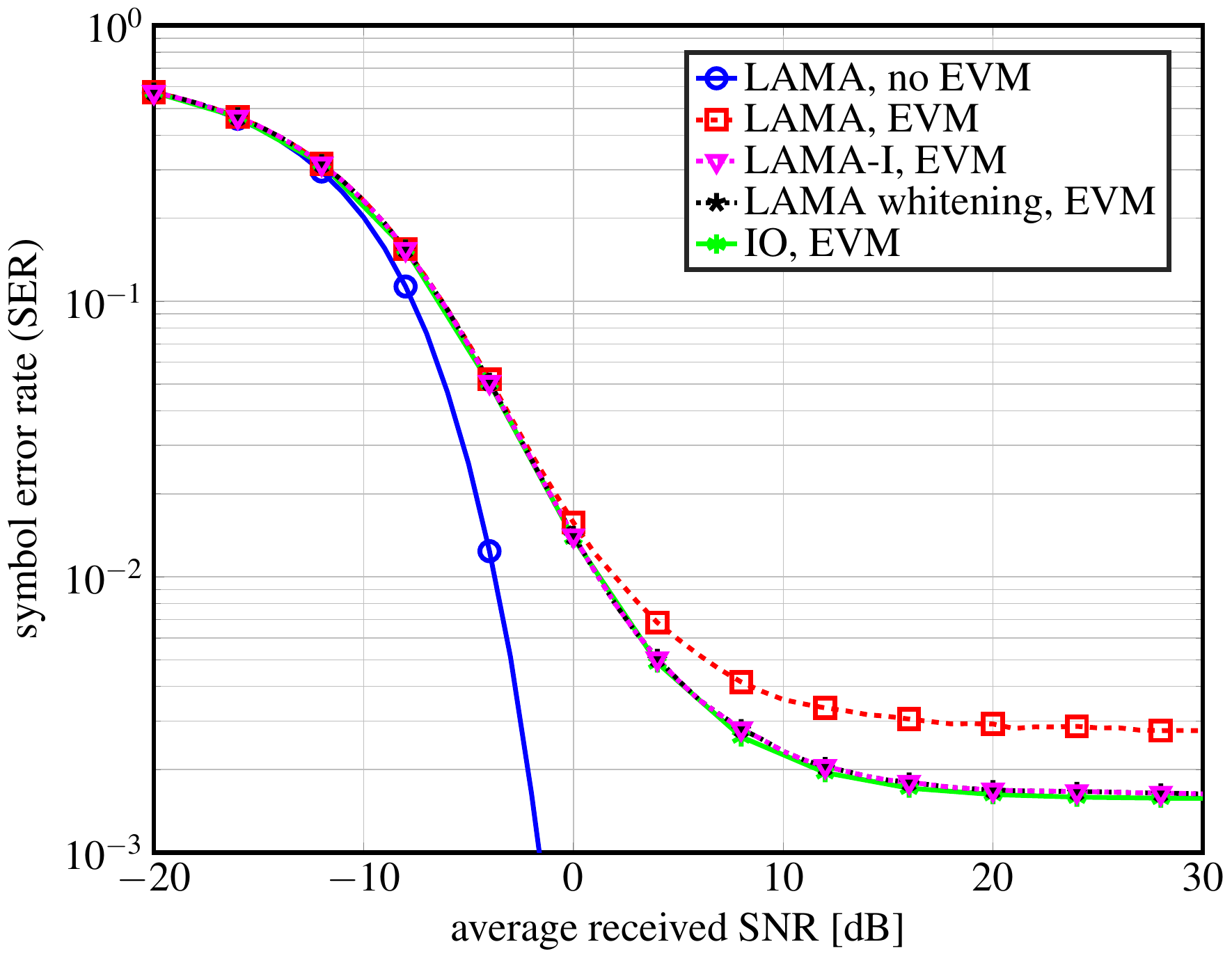}\label{fig:MIMO1}}
\hspace{0.2cm}
\subfigure[$128$ BS antennas and $128$ users.]{\includegraphics[width=0.98\columnwidth]{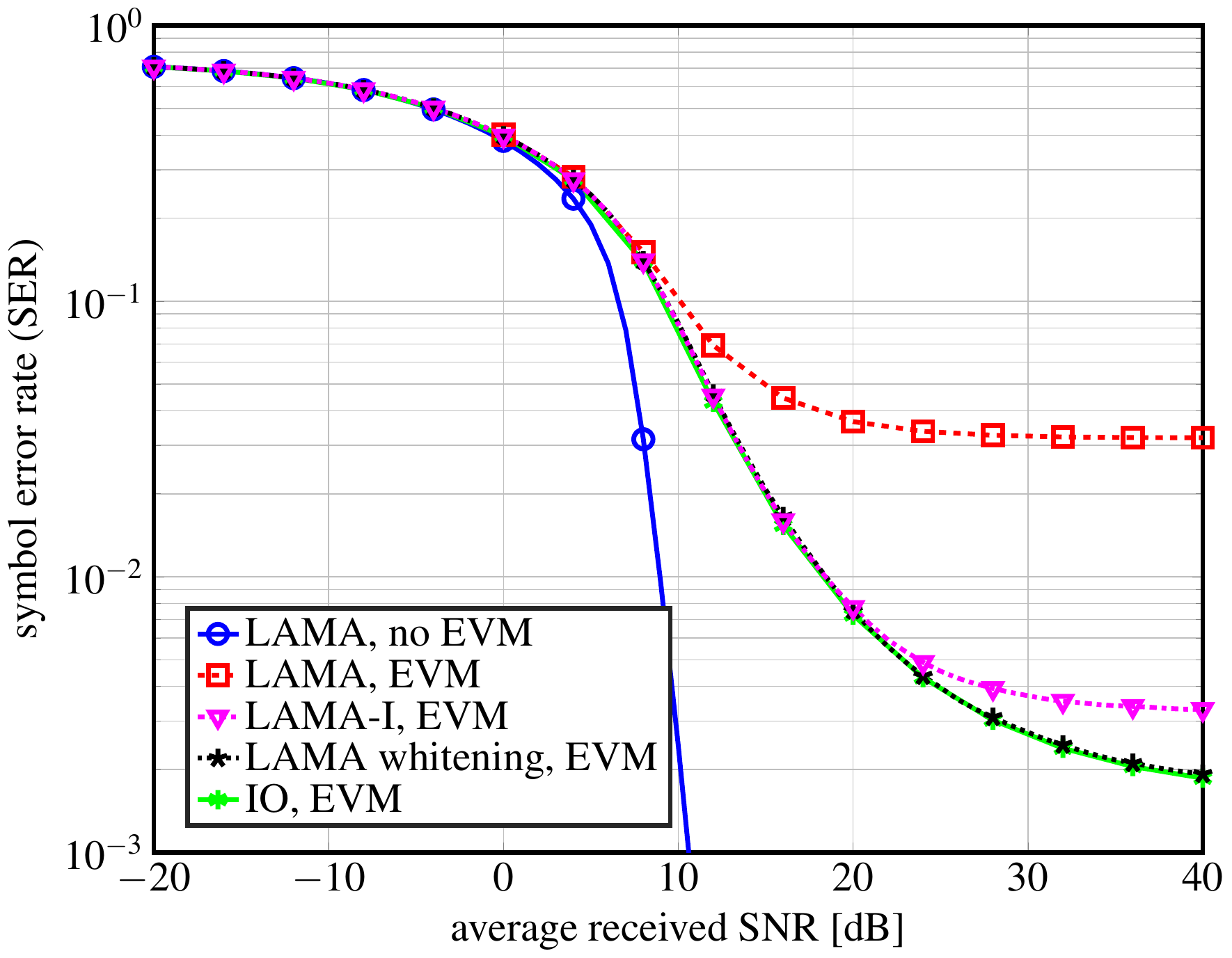}\label{fig:MIMO2}}	
\caption{Symbol error-rate of \mbox{LAMA-I} in large-MIMO systems with QPSK and strong Gaussian transmit noise ($\textit{EVM}= -10$\,dB). \mbox{\mbox{LAMA-I}} enables significant performance improvements compared to conventional LAMA and requires lower complexity than regular LAMA with noise whitening.}\label{fig:MIMO}
\end{figure*}
Figure~\ref{fig:MIMO} illustrates the performance of \mbox{LAMA-I} in a \mbox{$128 \times 8$} and \mbox{$128 \times 128$} large-MIMO system (we use the notation $\MR\times\MT$) with QPSK transmission, and transmit impairments modeled as i.i.d.\ circularly-symmetric complex Gaussian noise~\cite{Studer_Tx_OFDM}. 
We observe significant symbol error-rate (SER) improvements compared to that of regular LAMA, which achieves---given certain conditions on the MIMO system are met---the error-rate performance of the individually-optimal (IO) data detector in absence of transmit impairments (see~\cite{JGMS2015,jgms2015conf} for the details). 
We emphasize that \mbox{LAMA-I} entails virtually no complexity increase (compared to regular LAMA) and  achieves the same SER performance of whitening-based approaches, which require prohibitive computational complexity in large MIMO systems.




\subsection{Relevant Prior Art}

Channel capacity expressions for the transmit-impaired MIMO system model \eqref{eq:TNproblem} have first been derived in~\cite{Studer_Tx_OFDM}. A corresponding asymptotic analysis has been provided recently in~\cite{Tx_Replica}, which uses the replica method \cite{nishimori2001statistical} to obtain capacity expressions for large MIMO systems. The results in \cite{Studer_Tx_OFDM,Tx_Replica} build upon on the so-called  \emph{Gaussian transmit-noise model}, which assumes that  the transmit impairments in $\vece$ can be modeled as i.i.d.\ additive Gaussian noise that is independent of the transmit signal $\vecs$.
While the accuracy of this model for a particular RF implementation in a MIMO system using orthogonal frequency-division multiplexing (OFDM) has been confirmed via real-world measurements \cite{Studer_Tx_OFDM}, it may not be accurate for other RF transceiver designs and/or modulation schemes. 
\mbox{LAMA-I}, as proposed in this paper, enables us to study the fundamental performance of more general transmit impairments (which may, for example, exhibit statistical dependence with the transmit signal and even include deterministic non-linearities), which is in stark contrast to the commonly used transmit-noise model in~\cite{Studer_Tx_OFDM,Tx_Replica,studer2011system,schenk2008rf,schenk2005performance,goransson2008effect,suzuki2008transmitter,suzuki2009practical,gonzalez2011impact,gonzalez2011transmit,bjornson2013capacity,zhang2014mimo}. 
For the well-established Gaussian transmit-noise model, we will show in \fref{sec:transmitnoiseresults} that the state-evolution equations of \mbox{LAMA-I} coincide to the ``coupled fixed point equations'' in \cite{Tx_Replica}, which reveals that \mbox{LAMA-I} is a practical algorithm that delivers the same performance as predicted by replica-based capacity expressions in the large-system limit.

Data detection algorithms in the presence of transmit impairments were studied in~\cite{Studer_Tx_OFDM}. The proposed methods rely on the Gaussian transmit-noise model, which enables one to ``whiten'' the impaired system model \eqref{eq:TNproblem} by multiplying the received vector $\bmy$ with a so-called whitening matrix  $\bW=\No\bQ^{-\frac{1}{2}}$, where $\bQ=\NT\bH \bH^\Herm+\No\bI_M$ is the covariance matrix of the effective transmit and receive noise $\bmn+\bH\bme$, and $\NT$ denotes the variance of the entries of the transmit-noise vector $\bme$. By applying the whitening filter $\bW$ to the received vector in~\fref{eq:TNproblem}, we obtain the following statistically-equivalent, whitened input-output relation~\cite{Studer_Tx_OFDM,studer2011system}:
\begin{align} \label{eq:whitenedsystem}
\tilde{\bmy}=\widetilde{\bH}\bms+\tilde{\bmn},
\end{align}
where $\tilde\bmy=\bW\bmy$, $\widetilde\bH=\bW\bH$, and $\tilde\bmn=\bW(\bmn+\bH\bme)$. Optimal (as well as suboptimal) data detection can then be performed by considering  the whitened system model in  \eqref{eq:whitenedsystem}. 
While such a whitening approach enables optimal data detection in conventional, small-scale MIMO systems (see\cite{Studer_Tx_OFDM} for the details) under the Gaussian transmit-noise model, computation of the whitening matrix  $\bW$ quickly results in prohibitive computational complexity in large-MIMO systems consisting of hundreds of receive antennas---a situation that arises in massive MIMO~\cite{LETM2014,marzetta2010noncooperative,rusek2013scaling}, an emerging technology for 5G wireless systems. 
\mbox{LAMA-I} avoids computation of the whitening matrix~$\bW$ altogether, which results in (often significantly) reduced computational complexity. Furthermore, the generality of our system model enables \mbox{LAMA-I} to be resilient to a broader range of transmit-side impairments.

\subsection{Notation}

Lowercase and uppercase boldface letters designate column vectors and matrices, respectively. For a matrix $\bA$, we define its conjugate transpose to be $\bA^\Herm$. The entry on the $k$-th row and $\ell$-th column is $A_{k,\ell}$, and the $k$-th entry of a vector $\veca$ is~$a_k$. The $M\times M$ identity matrix is denoted by $\bI_M$ and the $M\times N$ all-zeros matrix by $\mathbf{0}_{M\times N}$. 
We denote the averaging operator by $\left\langle \bma \right\rangle = \frac{1}{N}\sum_{k=1}^N a_k$. Multivariate complex-valued Gaussian probability density functions (PDFs) are denoted by $\setC\setN(\bmm,\bK)$, where~$\bmm$ represents the mean vector and $\bK$ the covariance matrix; $\Exop_X\!\left[\cdot\right]$ and $\Varop_X\!\left[ \cdot\right] $ denote expectation and variance with respect to the PDF of the random variable~(RV)~$X$, respectively. We use $\mathbb{P}(X=x)$ to denote the probability of the RV $X$ being~$x$.

\subsection{Paper Outline}
The rest of the paper is organized as follows. \fref{sec:LAMA-I} details the \mbox{LAMA-I} algorithm along with the state-evolution framework. \fref{sec:Opt} provides conditions for which \mbox{LAMA-I} achieves the performance of the IO data detector. \fref{sec:transmitnoiseresults} analyzes the special case of Gaussian transmit-noise. We conclude in \fref{sec:conclusion}.


\section{LAMA-I: \underline{La}rge \underline{M}IMO \underline{A}pproximate Message Passing with Transmit \underline{I}mpairments}\label{sec:LAMA-I}

Large MIMO is believed to be one of the key technologies for 5G wireless systems~\cite{ABCHLAZ2014}. The main idea is to equip the base station (BS) with hundreds of antennas while serving a tens of users simultaneously and within the same frequency band. 
One of the key challenges in practical large MIMO systems is the high computational complexity associated with data detection  \cite{WBVSCJD2013}. 
We next introduce LAMA-I, a novel low-complexity data detection algorithm for large-MIMO systems that takes into account transmit-side impairments. We derive the associated complex state-evolution (cSE) framework, which will be used in Sections~\ref{sec:Opt} and~\ref{sec:transmitnoiseresults} to establish conditions for which LAMA-I achieves the error-rate performance of the IO data detector for the impaired system model \fref{eq:TNproblem}.

\subsection{Summary of the LAMA-I Algorithm}

In the remainder of the paper, we consider a complex-valued data vector $\vecs \in \complexset^{\MT}$, whose entries are chosen from a discrete constellation $\setO$, e.g., phase shift keying (PSK) or quadrature amplitude modulation (QAM). We further assume  i.i.d.\ priors $p(\vecs)=\prod_{\ell=1}^{\MT} p(s_{\ell})$ with
\begin{align} \label{eq:s_prior}
p\!\left(s_{\ell}\right) = \sum_{a\in \setO}p_a\delta\!\left(s_{\ell} - a\right)\!,
\end{align}
where $p_a$ corresponds to the (known) prior probability of the constellation point $a\in\setO$. In the case of uniformly distributed constellation points, we have $p_a=|\setO|^{-1}$, where $|\setO|$ is the cardinality of the set $\setO$.
We define the \emph{effective transmit signal} $\bmx=\bms+\bme$, which is distributed as $p(\vecx) = \prod_{\ell=1}^{\MT}p(x_{\ell})$ with 
\begin{align} \label{eq:lamaiprior}
p(x_{\ell}) = \int_\complexset p(x_{\ell} \vert s_{\ell}) p(s_{\ell}) \text{d}s_{\ell},
\end{align}
where $p(x_{\ell} \vert s_{\ell})$ models the transmit-side impairments. We can now rewrite the input-output relation \fref{eq:TNproblem}  as 
\begin{align} \label{eq:effectivesystem}
\bmy = \bH\bmx + \bmn.
\end{align}

\begin{figure}
\centering
\subfigure[Impaired MIMO system with LAMA-I as the data detector.]{\includegraphics[width=0.99\columnwidth]{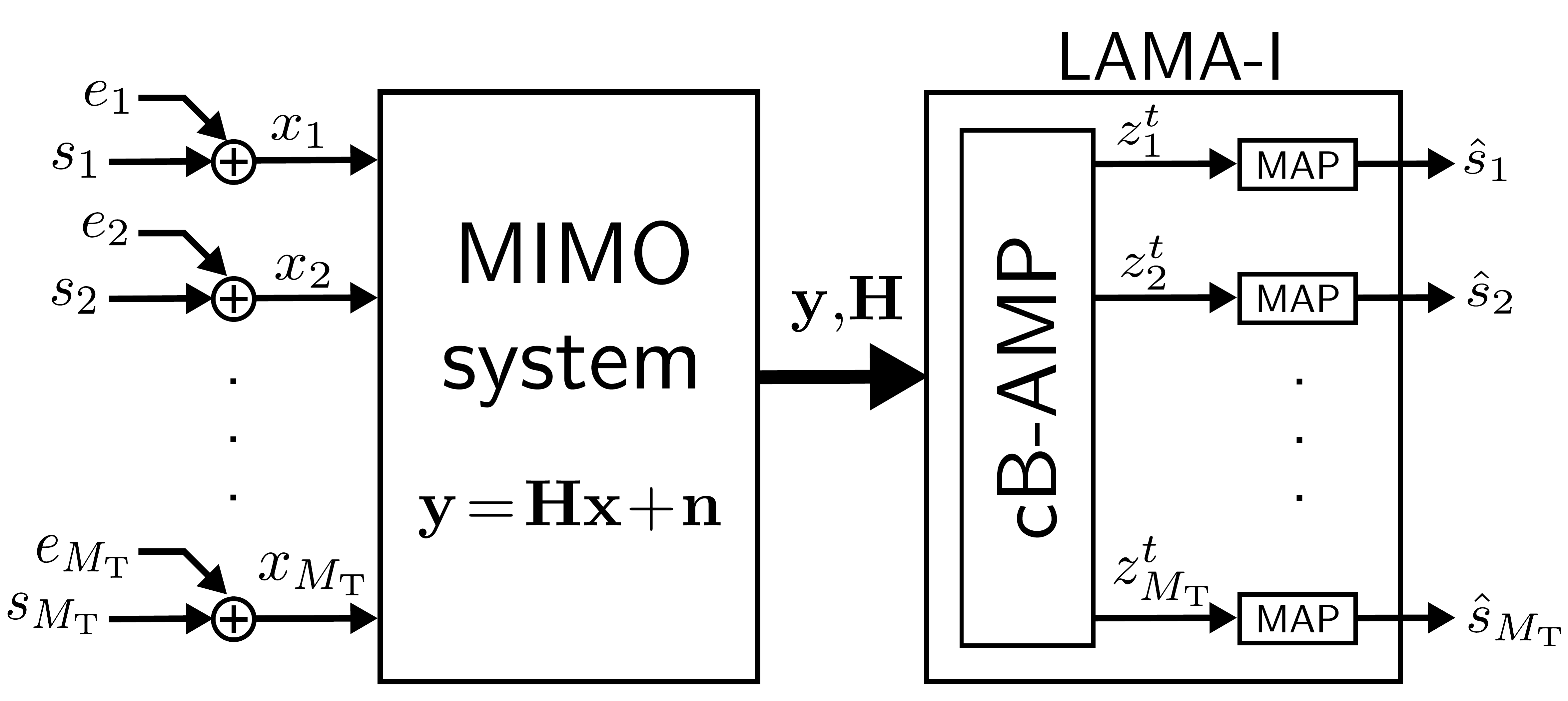}\label{fig:decouple_a}}
\subfigure[Equivalent decoupled MIMO system.]{\includegraphics[width=0.99\columnwidth]{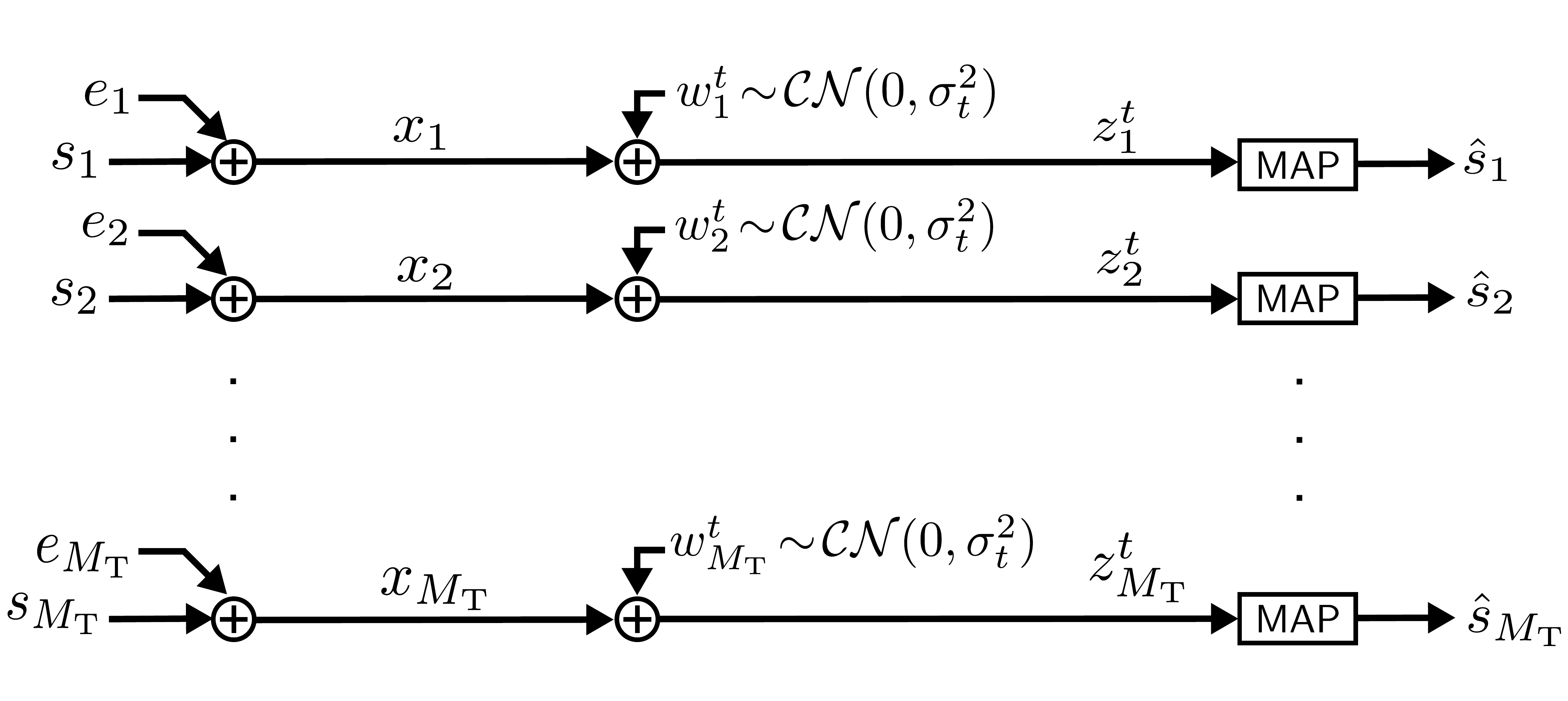}\label{fig:decouple_b}}
\caption{In the large-antenna limit, LAMA-I decouples the impaired MIMO system into $\MT$ parallel and independent AWGN channels, which allows us to perform impairment-aware MAP data detection, independently per user.} \label{fig:decouple}
\end{figure}

The key idea behind LAMA-I is to perform data detection in two steps.
We first use message passing on the factor graph for the distribution $p(\mathbf{s,x} | \mathbf{y, H})$ in order to obtain the marginal distribution $p(s_\ell | \mathbf{y, H})$. Once the message passing algorithm converged, we assume that it converged to the marginal distribution, which allows us to perform maximum a-posteriori (MAP) detection on $p(s_\ell | \mathbf{y, H})$ to obtain estimates~$\hat s_\ell$ for the transmit data signals \emph{independently} for every user. 
Since the factor graph for $p(\mathbf{s,x} | \mathbf{y, H})$ is dense, i.e., for every entry in the receive vector $\vecy$ we have a factor that is connected to every transmit symbol $x_\ell$, an exact message passing algorithm is computationally expensive. However, by exploiting the bipartite structure of the graph and the high dimensionality of the problem (i.e., both $\MT$ and $\MR$ are large), the entire algorithm can be simplified.\footnote{We refer to \cite{GJMS2015_TX} for more details on these claims.} 
In particular, we simplify our  message-passing algorithm using complex Bayesian AMP (cB-AMP) as proposed in \cite{JGMS2015,jgms2015conf} for the MIMO system model \fref{eq:effectivesystem}. cB-AMP calculates an estimate for the effective transmit signal~$\hat{x}_\ell$,~$\forall \ell$. 
The MAP estimate can then be calculated from $\hat{x}_\ell$ \emph{independently} for every user.  
The resulting two-step procedure of LAMA-I is illustrated in \fref{fig:decouple}.



As illustrated in \fref{fig:decouple_a}, we first use cB-AMP to compute the Gaussian output~$\vecz^t$ and the effective noise variance $\sigma^2_t$ at iteration $t$. Since the Gaussian output of cB-AMP can be modeled as $z_\ell^t=x_\ell+w_\ell^t$ with $w_\ell^t\sim\setN(0,\sigma^2_t)$, being independent from $x_\ell$, $s_\ell$ and $e_\ell$ $\forall \ell$, in the large-system limit (see \cite{andreaGMCS} for the details), the MIMO system is effectively decoupled into a set of~$\MT$ parallel and independent additive white Gaussian noise (AWGN) channels. \fref{fig:decouple_b} shows the equivalent decoupled system. Since the effective transmit signals are defined as $x_\ell=s_\ell+e_\ell$, $\ell=1,\ldots,\MT$, we have
\begin{align}\label{eq:equivalence}
z_\ell^t=s_\ell+e_\ell +w_\ell^t,
\end{align}
which allows us to compute the MAP estimate for each data symbol independently using
\begin{align}\label{eq:s_hat_l}
\hat{s}_\ell=\argmax_{s_\ell\in\setO}p(s_\ell | z_\ell^t).
\end{align}
Here, the probability $p(s_\ell | z_\ell^t)$ is obtained from Bayes' rule $p(s_\ell | z_\ell^t) \propto p(z_\ell^t | s_\ell) p(s_\ell)$ and from
\begin{align}\label{eq:MAP_posterior}
p(z_\ell^t | s_\ell) = \int_\complexset p(x_\ell=z_\ell^t-w_\ell^t | s_\ell) p(w_\ell^t) \dd w_\ell^t.
\end{align}

 




The resulting LAMA-I algorithm is summarized as follows. 
\newtheorem{alg}{Algorithm} 
\begin{alg}[\bf{LAMA-I}]\label{alg:LAMA-I} 
Initialize $\hat{x}^1_\ell=\Exop_X[X]$, \mbox{$\resid^1= \bmy$}, and $\tau^1=\beta\Varop_X[X]/\No$ with $\xrv \sim p(x_\ell)$ as defined in \fref{eq:lamaiprior}.
	\begin{enumerate}
	\item Run cB-AMP for $\tmax$ iterations by computing the following steps for $t=1,2,\ldots,\tmax$: 
	\begin{align*}
	\bmz^{t}&=\hat\bmx^t+\bH^\Herm\bmr^t\\
	\hat{\bmx}^{t+1} &=  \mathsf{F}(\bmz^{t},\No(1+\tau^t))\\
	\tau^{t+1} &=  \frac{\beta}{\No}\!\left\langle\mathsf{G}(\bmz^{t},\No(1+\tau^t))\right\rangle\\
	\resid^{t+1}  &=   \bmy-\bH\hat{\bmx}^{t+1}+\frac{\tau^{t+1}}{1+\tau^t}\resid^{t}.
	\end{align*}
	The scalar functions $\mathsf{F}(z^t_{\ell},\sigma_t^2)$ and $\mathsf{G}(z^t_{\ell},\sigma_t^2)$ operate element-wise on vectors, and correspond to the posterior mean and variance, respectively, defined as
	\begin{align}\label{eq:Ffunc}
	\!\!\!\! \mathsf{F}(z^t_{\ell},\sigma_t^2) &= 
	\int_\complexset x_{\ell}f(x_{\ell}\vert z^t_{\ell},\sigma_t^2)\mathrm{d}x_{\ell}\\\label{eq:Gfunc}
	\!\!\!\!  \mathsf{G}(z^t_\ell,\sigma_t^2)&=  \int_\complexset \abs{x_{\ell}}^2\!f(x_{\ell}\vert z^t_{\ell},\sigma_t^2)\mathrm{d}x_{\ell} \!-\! \abs{\mathsf{F}(z^t_{\ell},\sigma_t^2)}^2\!.
	\end{align}	
	Here, the message posterior distribution is $f(x_{\ell}\vert z^t_{\ell},\sigma_t^2)=\frac{1}{Z}p(z^t_{\ell}\vert x_{\ell},\sigma_t^2)p(x_\ell)$, where $p(z^t_{\ell}\vert x_{\ell},\sigma_t^2)\sim\setC\setN(x_{\ell},\sigma_t^2)$ and $Z$ is a normalization constant.
	\item Compute the MAP estimate using \fref{eq:s_hat_l} for $t=\tmax$ with the posterior PDF $p(s_\ell | z_\ell^\tmax)$ as defined in \fref{eq:MAP_posterior} and $p(w_\ell^\tmax) \sim \setN(0,\sigma^2_\tmax)$. The effective noise variance $\sigma^2_\tmax$ is estimated using the postulated output variance $\No(1+\tau^\tmax)$  from cB-AMP (see \cite[Def.~3]{JGMS2015}).
\end{enumerate}
\end{alg}
\subsection{State Evolution for LAMA-I}
Virtually all existing theoretical results are incapable of providing performance guarantees for the success of message-passing on dense graphs. In our specific application, however, the structure of the factor graph of $p(s_\ell,x_\ell | \mathbf{y, H} )$ enables us to study the associated theoretical properties in the large-system limit. 
As shown in \cite{JGMS2015}, the effective noise variance $\sigma_t^2$ can be calculated analytically for every iteration $t=1,2,\ldots,\tmax$, using the complex state evolution (cSE) recursion. 
In \fref{sec:Opt}, we will use the cSE framework to derive optimality conditions for which LAMA-I  achieves the error-rate performance of the IO data detector in \fref{eq:IOproblem}.
The cSE for cB-AMP is detailed in the following theorem.

\begin{thm}[{\!\!\cite[Thm.~3]{JGMS2015}}]\label{thm:CSE} Suppose that $p(\bmx)\sim\prod_{\ell=1}^\MT p(x_{\ell})$ and the entries of~$\bH$ are i.i.d.\ circularly-symmetric complex Gaussian with variance ${1}/{\MR}$. Let $\bmn\sim\setC\setN(\mathbf{0}_{\MR\times 1},\No\bI_\MR)$ and $\mathsf{F}$ be a pseudo-Lipschitz function as defined in \cite[Sec.~1.1, Eq.~1.5]{bayatimontanari}. Fix the system ratio $\beta=\MT/\MR$ and let $\MT\rightarrow\infty$. Then, the effective noise variance $\sigma^2_{t+1}$ of cB-AMP at iteration $t$  is given by the following cSE recursion:
	\begin{align}
	\sigma_{t+1}^2 &  
	= \No +\beta\Psi(\sigma_t^2). \label{eq:SErecursion}
	\end{align}
	Here, the mean-squared error (MSE) function $\Psi$ is defined by
	\begin{align}\label{eq:Psi}
	\Psi(\sigma_t^2) = \Exop_{\xrv,Z}\!\left[\abs{ \mathsf{F}\!\left(\xrv + \sigma_t Z,\sigma_t^2\right) - \xrv}^2 \right]
	\end{align}
	with $\xrv\sim p(x_\ell)$, $Z \sim\setC\setN(0,1)$, and $\mathsf{F}$ is defined in \fref{eq:Ffunc}. The cSE recursion  is initialized by $\sigma_1^2 = \No + \beta \Varop_X[X]$.
\end{thm}
The cSE in \fref{thm:CSE} tracks the effective noise variance $\sigma^2_t$ for every iteration $t$, which enables us to compute  the posterior distribution \fref{eq:MAP_posterior} required in Step 2) of \fref{alg:LAMA-I}.
\begin{rem} \label{rem:dependence}
The posterior mean function $\mathsf{F}$ and consequently the MSE function $\Psi(\sigma_t^2)$  in \fref{eq:Psi} depend on the effective transmit signal prior $p(\bmx)$ in \fref{eq:lamaiprior}, which is a function of the data-vector prior $p(\bms)$ and the conditional probability $p(\bmx|\bms)$ that models the transmit-side impairments. 
\end{rem}


\section{Optimality of \mbox{LAMA-I} } \label{sec:Opt}

We now analyze the optimality of LAMA-I for the impaired system model \fref{eq:TNproblem}.

\subsection{Optimality Questions}
The cSE framework enables us to characterize the performance of \mbox{LAMA-I} in the large-system limit. In this section, we use this framework to study the optimality of LAMA-I. In particular, we address the following two optimality questions: 
\begin{itemize}
\item[(i)] We derived \mbox{LAMA-I} using a message-passing algorithm. However, there exists a broader class of algorithms to accomplish the same task. More specifically, the version of \mbox{LAMA-I} that uses sum-product message passing employs the posterior mean function $\mathsf{F}$ as defined in \eqref{eq:Ffunc}. One can potentially change $\mathsf{F}$ (or even pick different functions at different iterations) and come up with estimates $\hat{x}_\ell$, $\forall \ell$, and perform data detection by using MAP detection on to these new estimates. Such alternative data-detection algorithms can still be analyzed through the state evolution framework. The first optimality question we can ask is whether we can improve the performance of \mbox{LAMA-I} by  choosing functions different to those we introduced in  \eqref{eq:Ffunc}? As we will show in Section \ref{sec:postmeanisoptimal}, the posterior mean functions we used in~\eqref{eq:Ffunc} are indeed optimal.  

\item[(ii)] We also ask ourselves whether the optimal \mbox{LAMA-I}, i.e., \mbox{LAMA-I} that sets $\mathsf{F}$ to a posterior mean, achieves the same error-rate performance as the IO data detector \fref{eq:IOproblem}?
\end{itemize}
In what follows, we will answer both of these questions in the large-system limit. 

\subsection{Optimality of Posterior Mean for \mbox{LAMA-I}}\label{sec:postmeanisoptimal}

Consider the following generalization of \mbox{LAMA-I}, where the posterior mean function is replaced with a general pseudo-Lipschitz function $\mathsf{F}_t$ \cite{andreaGMCS} that depends on the iteration step~$t$, i.e., where we use
\begin{align*}
	\hat{\bmx}^{t+1} &= \textstyle \mathsf{F}_t(\bmz^{t},\No(1+\tau^t)).
	\end{align*}
The first optimality question we would like to address is whether there exists a choice for the functions $\mathsf{F}_1, \mathsf{F}_2, \ldots$, such that the resulting data-detection algorithm achieves lower probability of error. The following theorem establishes the fact that it is impossible to improve upon the choice of \mbox{LAMA-I}, where we use the posterior mean. 


\begin{thm}\label{thm:Opt}
Let the assumptions made in \fref{thm:CSE} hold for $\mathsf{F}_1,\ldots,\mathsf{F}_\tmax$. Suppose that we run \mbox{LAMA-I} for $\tmax$ iterations and then, perform element-wise data detection. Let $\hat{s}_\ell$ be the estimate we obtain for $s_\ell$. We denote the detection error probability as $\mathbb{P}_{\mathsf{F}_1,\ldots,\mathsf{F}_\tmax} (\hat{s}_\ell \neq s_\ell)$ to emphasize on the dependence of this probability on the functions employed at every iteration. The choice of $\mathsf{F}_1,\ldots,\mathsf{F}_\tmax$ that minimizes $\mathbb{P}_{\mathsf{F}_1,\ldots,\mathsf{F}_\tmax} (\hat{s}_\ell \neq s_\ell)$ is the posterior mean employed in \eqref{eq:Ffunc}.
%
%
\end{thm}

A detailed version of the proof for this theorem is given in~\cite{GJMS2015_TX}. For the sake of brevity, we only sketch the main steps of the proof. Since $\mathsf{F}_1, \ldots, \mathsf{F}_{\tmax}$ are pseudo-Lipschitz according to Theorem \ref{thm:CSE}, we know that $z_\ell^{\tmax}$ can be modeled as 
\[
z_\ell^\tmax=s_\ell+e_\ell +w_\ell^\tmax,
\]
where $w_\ell^\tmax$ is Gaussian. The effect of $\mathsf{F}_1, \ldots, \mathsf{F}_{\tmax}$ is summarized by the variance of $w_\ell^\tmax$. It is straightforward to prove that the smaller the variance of $w_\ell^\tmax$ is, the smaller the error probability $\mathbb{P}_{\mathsf{F}_1,\ldots,\mathsf{F}_\tmax} (\hat{s}_\ell \neq s_\ell)$ will be. Hence, we should use a sequence of functions $\mathsf{F}_1, \ldots, \mathsf{F}_\tmax$ that minimize the variance of $w_\ell^\tmax$. We can use induction to establish that the posterior mean leads to the minimum variance. In the last iteration $\tmax$, if the variance of $w_\ell^{\tmax-1}$ is fixed, then it is straightforward to prove that we should use the posterior mean in the last iteration to minimize the variance of $w_\ell^\tmax$. By employing induction and by following the same line of argumentation, we can show that $\mathsf{F}_1, \ldots, \mathsf{F}_\tmax$ must all be the posterior mean. 

We now use the cSE framework in \fref{thm:CSE} to establish conditions for which \mbox{LAMA-I} is optimal. 
We consider the case where the number of iterations $\tmax \to \infty$ for which, as explained in \cite[Sec.~\uppercase\expandafter{\romannumeral4}]{JGMS2015}, the cSE recursion~\fref{eq:SErecursion} converges to the following fixed-point equation:
\begin{align}\label{eq:FixedPoint}
	\sigma^2 &  
	= \No +\beta\Psi(\sigma^2).
\end{align}
This equation can in general have one or more fixed points. If it has more than one fixed point, then  \mbox{LAMA-I} may converge to different fixed points, depending on its initialization~\cite{ZMWL2015}.

As the first step toward proving that \mbox{LAMA-I} is optimal, we derive conditions under which the fixed point equation \eqref{eq:FixedPoint} has a unique solution. To establish such conditions, we first define the following quantities~(also see \cite[Defs.~1-4]{jgms2015conf}).
\begin{defi}\label{def:betaN0}
For a given transmit data-vector prior $p(\vecs)$ and transmit-impairment distribution $p(\bmx|\bms)$, we define the exact recovery threshold (ERT) $\betamax$ and the minimum recovery threshold (MRT) $\betamin$ as
\begin{align*}
\betamax= \min_{\sigma^2>0}\!\left\{\!\left(\frac{\Psi(\sigma^2)}{\sigma^2}\right)^{\!\!-1}\right\}\!, \,
\betamin&=\min_{\sigma^2>0}\!\left\{\!\left(\frac{\textnormal{d}\Psi(\sigma^2)}{\textnormal{d}\sigma^2}\right)^{\!\!-1}\right\}\!.
\end{align*}
The minimum critical noise $\Nomin$ is defined as 
\begin{align*}
\Nomin &=  \min_{\sigma^2>0} \!\left\{\sigma^2-\beta\Psi(\sigma^2):\beta\frac{\textnormal{d}\Psi(\sigma^2)}{\textnormal{d}\sigma^2}=1\right\}\!,
\end{align*}
and the maximum guaranteed noise $\Nomax$ is defined as  
\begin{align*}
\Nomax&= \max_{\sigma^2>0}\!\left\{\sigma^2 - \beta\Psi(\sigma^2):\beta\frac{\textnormal{d}\Psi(\sigma^2)}{\textnormal{d}\sigma^2} = 1\right\}\!.
\end{align*}
\end{defi}	

Using \fref{def:betaN0}, the following theorem establishes several regimes in which the fixed point of \mbox{LAMA-I} is unique.
\begin{lem}[Optimality Conditions of \mbox{LAMA-I}]\label{lem:betaN0}
Let the assumptions made in \fref{thm:CSE} hold and let $\tmax\to\infty$. Fix $p(\vecs)$ and $p(\vecx|\vecs)$.  If the variance of the receive noise $\No$ and system ratio $\beta$ are in one of the following three regimes:
\begin{enumerate}
\item $\beta \in \left(0,\betamin\right]$ and $\No \in \reals^+$
\item $\beta \in \left(\betamin,\betamax\right)$ and $\No \in \left[0,\Nomin\right) \cup \left(\Nomax,\infty\right)$
\item $\beta \in \left[\betamax,\infty\right)$ and $\No \in \left(\Nomax,\infty\right)$
\end{enumerate}
then \mbox{LAMA-I} solves the optimal problem.
\end{lem}

The proof follows from \cite[Table \uppercase\expandafter{\romannumeral 2}]{jgms2015conf}. Note that for \mbox{LAMA-I}, the quantities in \fref{def:betaN0} do not only depend on the data-vector prior $p(\vecs)$, but also on the transmit-impairment distribution $p(\vecx|\vecs)$ (cf.~\fref{rem:dependence}). 

%

\subsection{\mbox{LAMA-I} vs. Individually Optimal (IO) Data Detection}
	
We now show that in the large-system limit, \mbox{LAMA-I} achieves the error-rate performance of the IO data detector~\fref{eq:IOproblem}, if the fixed-point equation \fref{eq:FixedPoint} has a unique fixed point. As will be clear from our arguments, even in cases where \mbox{LAMA-I} does not have a unique fixed point, one of its fixed points corresponds to the solution of IO. It is, however, difficult to find a suitable algorithm initialization that would cause our method to converge to the optimal fixed point. The core of our optimality analysis is the result on the performance of IO data detection based on the replica analysis presented in \cite{GV2005}. The replica analysis for IO data detection makes the following assumption about~$\hat{s}^\Opt_\ell$.

\begin{defi}\label{def:hardsoft}
The IO solution is said to satisfy hard-soft assumption, if and only if there exist a function $D: \mathbb{R} \rightarrow \mathcal{O}$, whose set of discontinuities has Lebesgue measure zero and
\[
\hat{s}^\Opt_\ell = D(\mathbb{E} (s_\ell | \mathbf{y, H}) ).
\] 
\end{defi}


For some popular constellation sets, we can prove that the hard-soft assumption is in fact true. For example, for equiprobable BPSK constellation points, we have
\[
\mathbb{E} ({s}_\ell | \mathbf{y, H}) = \mathbb{P} ({s}_\ell = 1 | \mathbf{y, H}) - \mathbb{P} ({s}_\ell = -1 | \mathbf{y, H}),
\] 
and hence, $\hat{s}^\Opt_\ell  = \sign(\mathbb{E} ({s}_\ell | \mathbf{y, H})) $.

The next theorem establishes conditions for which LAMA-I achieves  the performance of the IO data detector. 

\begin{thm} \label{thm:IOptimality}
Suppose that the IO solution satisfies the hard-soft assumption. Furthermore, assume that the assumptions underlying  the replica symmetry in \cite{GV2005} are correct. Then, under all the conditions of Lemma \ref{lem:betaN0} and in the large-system limit, the error probability of \mbox{LAMA-I} is the same as probability of error of the IO data detector.  
\end{thm}
For the sake of brevity, we only present a proof sketch; see~\cite{GJMS2015_TX} for the proof details.
From the hard-soft assumption we realize that in order to characterize the probability of error of the IO data detector, we have to characterize the joint distribution of $(s_\ell, \mathbb{E} (s_\ell | \mathbf{y, H}))$. Note that in   \cite{GV2005} the limiting distribution of $(x_\ell, \mathbb{E} (x_\ell | \mathbf{y, H}))$ is calculated. A similar approach will work for our problem too. However, we have to slightly modify the problem and make it closer to the one in~\cite{GV2005}. As the first step, we first derive the limiting distribution of $(s_\ell, x_\ell, \mathbb{E} (s_\ell | \mathbf{y,H}))$. Note that the joint distribution of $(s_\ell, x_\ell)$ is known. Furthermore, $s_\ell \to x_\ell \to \mathbf{y}$ form a Markov chain. Hence, $s_\ell \to x_\ell \to \mathbf{E} (x_\ell | \mathbf{y,H})$ is a Markov chain, and conditioned on $x_\ell$, the two quantities $s_\ell$ and $\mathbf{E} (x_\ell | \mathbf{y,H})$ are independent. This implies that in order to characterize the distribution of $(s_\ell, x_\ell, \mathbb{E} (s_\ell | \bmy, \bH))$, we only need to characterize the distribution of $(x_\ell, \mathbb{E} (s_\ell | \mathbf{y,H}))$. Furthermore, we have
\begin{align*}
\mathbb{E} (s_\ell | \bmy,\bH) = \int \mathbb{E} (s_\ell | x_\ell) \dd p(x_\ell | \bmy,\bH). 
\end{align*}
Define $L(x_\ell) = \mathbb{E} (s_\ell | x_\ell)$. Our original problem of characterizing the limiting distribution of $(s_\ell, \mathbb{E} (s_\ell | \mathbf{y, H}))$ is simplified to characterizing the limiting distribution of $(x_\ell, \mathbb{E} (L(x_\ell)| \mathbf{y,H}))$. This latter problem can be solved by the replica method as explained in \cite{nishimori2001statistical}. The final result is the following: the joint distribution of  $(s_\ell, x_\ell, \mathbb{E} (s_\ell | \mathbf{y,H}))$ converges to $(S, X, \mathbb{E} (S | X+ \tilde{\sigma} Z))$, where $S \sim p(s_\ell)$, $X | S=s \sim p(x_\ell | s_\ell )$, $Z \sim N(0,1)$ and is independent of both $S$ and $X$, and finally~$\tilde{\sigma}$ satisfies the fixed point equation 
\begin{eqnarray}\label{eq:fpreplica}
\tilde{\sigma}^2 = N_0+ \beta \Psi(\tilde{\sigma}^2).
\end{eqnarray}
Note that this is the same fixed point equation as the one we have for LAMA-I~\fref{eq:FixedPoint}. Hence, whenever \eqref{eq:fpreplica} has a unique fixed point, the replica analysis and LAMA-I will necessarily lead to the same solution. So far, we have shown that the effective noise level is the same for \mbox{LAMA-I} and IO. It is straightforward to show that since the effective noise levels are the same, the error probability of both schemes is the same. For the details, refer to our journal paper \cite{GJMS2015_TX}. 



\section{LAMA-I for the Gaussian Transmit-Noise~Model}
\label{sec:transmitnoiseresults}

\fref{thm:Opt},  \fref{lem:betaN0}, and \fref{thm:IOptimality} as given above hold for general transmit-impairment distributions $p(\vecx|\vecs)$.
We now focus on the well-established Gaussian transmit-noise model~\cite{Studer_Tx_OFDM,schenk2008rf}. In particular, we start by providing the remaining \mbox{LAMA-I} algorithm details and then, derive more specific conditions for which LAMA-I is optimal. We furthermore provide simulation results for  finite-dimensional systems. 

\subsection{Algorithm Details}

We assume $\bme \sim \setC \setN (0,\NT\bI_{\MT})$, where $\bme$ is  independent from $\vecs$ and $\vecn$, and $\NT$ is the transmit-noise power. 
The following lemma provides the remaining details for  \fref{alg:LAMA-I} with this model. The proof is given in \fref{app:FG_MIMO}.

\begin{lem}
Assume the MIMO system in \fref{eq:TNproblem} with $\bme \sim \setC \setN (0,\NT\bI_{\MT})$ being  independent of $\bms$ and~$\vecn$. For Step 1) of \fref{alg:LAMA-I}, the probability distribution $p(x_\ell)$ is given by 
\begin{align*}
p(x_\ell) &= \sum_{a\in \setO}p_a \frac{1}{\pi\NT}\exp\!\left(-\frac{1}{\NT}|x_\ell-a|^2\right)\!.
\end{align*}
The posterior mean $\mathsf{F}$ and variance $\mathsf{G}$ function corresponds to
\begin{align*}\nonumber
\mathsf{F}(z^t_\ell,\sigma_t^2)&= \frac{\NT}{\NT+\sigma_t^2}z^t_\ell + \frac{\sigma_t^2}{\NT+\sigma_t^2} \sum_{a\in \setO} w_a a,\\
\mathsf{G}(z^t_\ell,\sigma_t^2)&= \frac{\NT\sigma_t^2}{\NT+\sigma_t^2} + \sum_{a\in \setO} w_a \! \left|\frac{\NT z^t_\ell+\sigma_t^2 a}{\NT+\sigma_t^2} - \mathsf{F}(z^t_\ell,\sigma_t^2) \right|^2\!\!,
\end{align*}	
respectively, with
\begin{align}\label{eq:w_a}
w_a=\frac{p_a \exp\left( -\frac{|z^t_\ell-a|^2}{\NT+\sigma_t^2}\right)}{\sum\limits_{a\in \setO} p_a \exp\left( -\frac{|z^t_\ell-a|^2}{\NT+\sigma_t^2}\right)}.
\end{align}
For Step 2), the MAP estimator \fref{eq:s_hat_l} is given by 
\begin{align} \label{eq:inputnoiseMAP}
\hat{s}_\ell=\argmin \limits_{a\in\setO} \left(\frac{|z_\ell^\tmax-a|^2}{\NT+\No(1+\tau^\tmax)}-\log p_a\right)\!.
\end{align}
%
\end{lem}

For the Gaussian transmit-noise model, we see that \mbox{LAMA-I} only requires a few subtle modifications to the functions $\mathsf{F}$ and $\mathsf{G}$ compared  to regular LAMA \cite[Alg.~1]{JGMS2015}, which ignores transmit-side impairments. Hence, making LAMA robust to the Gaussian transmit-noise impairments comes at virtually no expense in terms of complexity, but results in often significant performance improvements (cf.~\fref{sec:simulations}). 

\subsection{Optimality Conditions}

The optimality conditions in \fref{lem:betaN0}, which depend on the system ratio $\beta$, receive noise variance $\No$, as well as the signal prior and the transmit-impairment model, can be obtained via the fixed-point equation in \eqref{eq:FixedPoint}. 
%

It can be shown that for the Gaussian transmit noise model, the fixed-point equation~\eqref{eq:FixedPoint} is equivalent to the ``coupled fixed point equations'' derived in \cite[Eqs.~48 and~49]{Tx_Replica}, which have been used to characterize the capacity of the impaired system~\fref{eq:TNproblem}. While the results in \cite{Tx_Replica} have been obtained via the replica method \cite{nishimori2001statistical}, LAMA-I provides a practical algorithm that achieves the same performance in the large-system limit. 

The following lemma provides a condition for which \mbox{LAMA-I} is optimal. Our condition is independent of the receive noise variance $\No$ and the transmit-noise power $\NT$. The proof is given in \fref{app:betamin}.
\begin{lem}\label{lem:betamin}
Let the assumptions in \fref{thm:CSE} hold and suppose that the IO solution satisfies the hard-soft assumption. Define $\beta^{\textnormal{min}}_{\textnormal{m}}=\min_{\NT}\betamin(\NT)$. Furthermore, assume the Gaussian transmit-noise model. If $\beta \leq \beta^{\textnormal{min}}_{\textnormal{m}}$, then LAMA-I is optimal.
\end{lem}

This lemma implies that there is a threshold $\beta^{\textnormal{min}}_{\textnormal{m}}$ on the system ratio $\beta$ that enables LAMA-I to achieve the same error-rate performance as the IO data detector in the large-system limit. Note that this condition is independent of the receive and transmit noise levels $\No$ and $\NT$, respectively. 
%


\subsection{Simulation Results}\label{sec:simulations}

We now demonstrate the efficacy of LAMA-I for the Gaussian transmit-noise model in more realistic, finite-dimensional large-MIMO systems. We define the average receive signal-to-noise-ratio (SNR) as 
 \begin{align*}
\SNR=\frac{\Exop\left[\|\bH \vecs\|^2\right]}{\Exop\left[ \|\vecn\|^2\right]}=\beta \frac{\Es}{\No},
 \end{align*}
where $\Es=\Exop\left[ |s_{\ell}|^2 \right]$. We also define the so-called error-vector magnitude (EVM) as 
\begin{align*}
\textit{EVM}=\frac{\Exop\left[ \|\vece\|^2 \right]}{\Exop\left[ \|\vecs\|^2 \right]}=\frac{\NT}{\Es.}
\end{align*}

Figures~\ref{fig:MIMO1} and \ref{fig:MIMO2} illustrate the symbol error rate (SER) simulation results for large MIMO systems  with QPSK modulation, Gaussian transmit-noise, and two antenna configurations, i.e., $128 \times 8$ and $128 \times 128$. In both figures, the solid blue line corresponds to the performance of regular LAMA \cite{JGMS2015,jgms2015conf} in absence of transmit noise (i.e., $\textit{EVM} = -\infty$\,dB). As shown by the dashed red line, regular LAMA experiences a significant performance loss in the presence of transmit noise with $\textit{EVM}=-10$\,dB. In contrast, LAMA-I (indicated with the dash-dotted magenta line) yields significant performance improvements (the maximum number of LAMA-I iterations for $128\times8$ and $128\times128$ was  $\tmax=10$ and $\tmax=15$, respectively). The solid green line shows the optimal large-system limit performance. As it can be seen, LAMA-I closely approaches the optimum SER performance for finite-dimensional systems. 

Figures~\ref{fig:MIMO1} and \ref{fig:MIMO2}  furthermore compare LAMA-I to regular LAMA operating on the whitened system \fref{eq:whitenedsystem} shown by the dotted black line.
While both approaches achieve near-optimal performance, the whitening-based approach entails prohibitive complexity, mainly caused by the inverse matrix square root. In addition, the whitening-based approach is designed specifically for the Gaussian transmit-noise model; in contrary, LAMA-I is applicable to a broader range of real-world transmit-side  impairments.


\section{Conclusion}
\label{sec:conclusion}

We have introduced LAMA-I, a novel, computationally efficient data detection algorithm suitable for large-MIMO systems that are affected by a broad range of transmit-side impairments. We have developed conditions in the large-system limit for which LAMA-I achieves the error rate performance of the individually optimal (IO) data detector. For the special case of the Gaussian transmit-noise model and for practical antenna configurations, we have demonstrated that LAMA-I enables significant performance improvements compared to impairment-agnostic algorithms at virtually no overhead in terms of computational complexity.
As a consequence, \mbox{LAMA-I} is a practical data-detection algorithm that renders practical large-MIMO systems more resilient to user equipment that suffers from strong transmit-side impairments.

%

\appendices

\section{Derivation of $\mathsf{F}$ and $\mathsf{G}$ for Gaussian Transmit Noise}\label{app:FG_MIMO}

We first derive $p(\bmx)$ as used in Step 1) of~\fref{alg:LAMA-I}. From~\fref{eq:s_prior} and the Gaussian transmit-noise model $\bme \sim \setC \setN (0,\NT\bI_{\MT})$, which assumes independence from $s_\ell$, we can write effective transmit signal prior \fref{eq:lamaiprior} as follows:
\begin{align*}
	p(x_\ell)
	&= \int_\complexset\frac{1}{\pi\NT}\exp\!\left(-\frac{1}{\NT}|s_\ell-x_\ell|^2\right) \sum_{a\in \setO}p_a\delta\!\left(s_\ell - a\right)\! \dd s \\
	&= \sum_{a\in \setO}p_a \frac{1}{\pi\NT}\exp\!\left(-\frac{1}{\NT}|x_\ell-a|^2\right)\!.
\end{align*}
With this result, we can write the message posterior distribution $f(x_\ell\vert\hat{x}_\ell,\sigma_t^2)$ defined in Step 1) of \fref{alg:LAMA-I} as:
\begin{align*}
	f(x_\ell\vert z^t_\ell,\sigma_t^2)
	&=\frac{1}{Z\pi^2\NT\sigma_t^2}\sum_{a\in \setO} p_a \exp\!\left( -\frac{|z^t_\ell-a|^2}{\NT+\sigma_t^2}\right) \\
	&\times \exp\!\left( -\frac{\NT+\sigma_t^2}{\NT\sigma_t^2} \left|x_\ell-\frac{\NT z^t_\ell+\sigma_t^2 a}{\NT+\sigma_t^2} \right| ^2\right).
\end{align*}
Here the normalization constant $Z$ is chosen so that $\int_{\complexset}f(x_\ell\vert\hat{x}_\ell,\sigma_t^2)\dd x_\ell=1$, which can be computed as:
\begin{align*}
	Z=\sum_{a\in \setO} p_a \frac{1}{\pi(\NT+\sigma_t^2)} \exp\left( -\frac{|z^t_\ell-a|^2}{\NT+\sigma_t^2}\right)\!.
\end{align*}
Therefore, the posterior mean $\mathsf{F}(z^t_\ell,\sigma_t^2)$ in \fref{eq:Ffunc} is given by:
\begin{align*}
	\mathsf{F}(z^t_\ell,\sigma_t^2)=\int_{\complexset}x_\ell f(x_\ell\vert z^t_\ell,\sigma_t^2) \dd x_\ell=\sum_{a\in \setO} w_a \frac{\NT z^t_\ell+\sigma_t^2 a}{\NT+\sigma_t^2},
\end{align*}
with the shorthand notation \fref{eq:w_a}.
The message variance $\mathsf{G}(z^t_\ell,\sigma_t^2)$ defined in \fref{eq:Gfunc} can be derived similarly.
%
%

For Step 2) of \fref{alg:LAMA-I}, the effective noise $w_\ell^\tmax$ is distributed as $p(w_\ell^\tmax) \sim \setC\setN\left(0,\No(1+\tau^\tmax)\right)$ with statistical independence from the transmit-noise model $e_\ell \sim \setC \setN(0,\NT)$, which yields $p(w_\ell^\tmax+e_\ell) \sim \setC\setN\left(0,\No(1+\tau^\tmax)+\NT\right)$. 
With this result and relation \fref{eq:equivalence}, we have $p(z_\ell^\tmax|s_\ell) \sim \setC \setN\left(s_\ell,\No(1+\tau^\tmax)+\NT\right)$, which together with \fref{eq:s_prior}, yields the following posterior distribution:
\vspace{0.05cm}
\begin{align}\nonumber
	p(s_\ell|z_\ell^\tmax) \propto 
	& \,\, p(s_\ell) p(z_\ell^\tmax|s_\ell) \\[0.1cm]
	\nonumber
	= & \sum\limits_{a \in \setO}  \delta\!\left(s_{\ell} - a\right) \frac{p_a}{\pi (\NT+\No(1+\tau^\tmax))} \\
	& \times \exp\!\left( \frac{-|z_\ell^\tmax-s_\ell|^2}{\NT+\No(1+\tau^\tmax)}\right)\!.\label{eq:posterior}
\end{align}
Using the posterior distribution given in \fref{eq:posterior}, we now compute the MAP estimator \fref{eq:s_hat_l}:
\begin{align*}
	\hat{s}_\ell
&=\argmax_{s_\ell\in\setO} \sum\limits_{a \in \setO} \delta\!\left(s_{\ell} - a\right) p_a \exp\!\left( \frac{-|z_\ell^\tmax-s_\ell|^2}{\NT+\No(1+\tau^\tmax)}\right)\\
&=\argmin \limits_{a\in\setO}\!\left(\frac{|z_\ell^\tmax-a|^2}{\NT+\No(1+\tau^\tmax)}-\log p_a\right)\! .
\end{align*}

\section{Proof of \fref{lem:betamin}}\label{app:betamin}
If $\beta \leq \min_{\NT}\betamin(\NT)$, then  $\beta \leq \beta^\textnormal{min}_\textnormal{m} \leq \betamin(\NT)$ for any $\NT$. As a result, by \fref{lem:betaN0}, LAMA-I achieves the performance of the IO problem \fref{eq:IOproblem} for any $\No$ and $\NT$.

\vspace{0.3cm}
\IEEEtriggeratref{10}

\bibliographystyle{IEEEtran}
\bibliography{bib/VIPconfs-jrnls,bib/publishers,bib/VIP_07_05_2015}

\end{document}